# Electronically Rather Than Geometrically Driven Concerted Proton Transfer under Graphene Confinement

Le Jin,[1] Depeng Zhang,[1] Xinrui Yang,[2] Xiaoan Li,[3,4] Bo Wang,[5] Yang Gao[6,a)]

## AFFILIATIONS

[1] *School of Science, Shenyang University, Shenyang 110044, China*

[2] *Key Laboratory of Theoretical and Computational Photochemistry, Ministry of Education, College of Chemistry, Beijing Normal University, Beijing 100875, China*

[3] *NHC Key Laboratory of Nuclear Technology Medical Transformation, Mianyang Central Hospital, School of Medicine, University of Electronic Science and Technology of China, Mianyang, Sichuan, 621099, China*

[4] *Department of Gastroenterology, National Clinical Key Specialty, Mianyang Central Hospital, School of Medicine, University of Electronic Science and Technology of China, Mianyang 621000, China*

[5] *School of Science, Northeast Electric Power University, Jilin 132012, China*

[6] *School of Medicine, Southwest University of Science and Technology, Mianyang 621010, China*

[a)] Author to whom correspondence should be addressed: ygaoxs@gmail.com

**ABSTRACT:** Proton transfer in water is a common yet mechanistically non-trivial phenomenon that underpins proton conduction and chemical reactivity. However, how confinement alters it, and whether the changes are driven by geometry or by electronic interaction, remains far from fully understood. Here, using a water hexamer confined between two graphene sheets as a model system, we combine coupled-cluster and first-principles calculations with energy decomposition to investigate how confinement alters its concerted proton transfer. We find that tightening the confinement lowers the electronic proton-transfer barrier. This reduction does not arise from the geometric distortion of the ring, which merely accompanies the confinement, but from the orbital stabilization of the transition state by the graphene walls. These results demonstrate that the proton-transfer barrier under confinement is governed by electronic rather than geometric factors, providing a mechanistic design principle for tailoring proton conduction and chemical reactivity through nanoscale confinement.

Water under nanoscale confinement departs markedly from the bulk, exhibiting anomalously low dielectric constant, superionic proton conductivity, and restructured hydrogen-bond networks.[1-5] Proton transfer underpins proton conduction, from the superionic state of confined water to nanofluidic devices and proton-exchange membranes.[6-9] In one-dimensional carbon nanotubes the excess proton migrates through a distorted Zundel configuration far faster than in bulk water,[10, 11] and in two-dimensional graphene slits it is predominantly solvated as an Eigen cation with altered transfer dynamics.[12] Scanning-probe microscopy has directly imaged the Eigen and Zundel cations and their interconversion in monolayer water on metal surfaces, a process that entails interfacial proton transfer.[13] Subnanometer confinement further drives water toward a molecular superionic state in which protons migrate through concerted chain-like transfers with low barriers.[14-16] Despite this progress, these studies mainly address proton transport. Even the studies that examined confinement effects on proton-transfer energetics concerned a single hydronium-to-water transfer[17] or water self-dissociation,[18] rather than the concerted transfer of a neutral cyclic water cluster. The water hexamer is an ideal model for this process, because its closed hydrogen-bond network makes each molecule both donor and acceptor, and proton transfer proceeds through a cyclic transition state where all six O-H bonds stretch and contract in concert.[19-22] The six oxygen atoms of the hexamer lie nearly in a plane, and its transfer occurs in that plane. The planar six-membered ring is in fact a stable motif on carbon substrates, where such rings self-assemble into a two-dimensional ice on graphite.[23] A two-dimensional graphene slit, whose interlayer separation can be tuned so that the interaction ranges from weak attraction to strong repulsion, therefore offers a natural geometry for probing how spatial constraint affects proton transfer.[24-26] The geometric and electronic contributions to the proton-transfer barrier under confinement have not been disentangled. Earlier first-principles work attributed the lowering of water rotation barriers in a carbon nanotube to Pauli repulsion.[27] Rotation, however, is a non-covalent reorientation, whereas proton transfer breaks and forms covalent bonds, so whether this Pauli-repulsion picture extends to proton transfer remains an open question.

In this work, we combine DLPNO-CCSD(T) barrier calculations with sobEDA energy decomposition and the activation strain model to disentangle the geometric and electronic contributions to the barrier. We find that the barrier reduction does not arise from the geometric changes, which accompany but do not cause it. It instead originates from the electronic interaction between the water ring and the graphene walls. Our results thus identify orbital interaction, rather than geometric changes, as the factor that controls concerted proton transfer under graphene confinement. This mechanism is resolved in detail below.

The confinement model consists of two graphene sheets with dimensions of 13 Å × 14 Å. This size balances computational cost and the need to minimize boundary effects, being large enough to accommodate the water hexamer with sufficient clearance from the sheet edges and small enough to enable the calculations. The edge dangling bonds were passivated with hydrogen atoms to avoid boundary effects, and the graphene atoms were kept fixed during all relaxations and transition-state searches to maintain a well-defined confinement environment. A cyclic water hexamer was placed between the two sheets, and the interlayer distance was varied from 7.0 Å to 4.5 Å in six steps. At 7.0 Å the water molecules and the graphene walls are separated beyond their equilibrium van der Waals distance, in the weakly attractive regime, whereas at 4.5 Å the confinement is so tight that the electron clouds of the water and graphene begin to overlap, giving rise to strong repulsion. The intermediate spacings thus span the continuous transition from attraction to repulsion, allowing a systematic investigation of how the confinement strength affects the structure and proton transfer of the water hexamer. The upper limit was set at 7.0 Å because at larger spacings the angle between the water-ring and graphene planes increases sharply and the transition state drifts toward the graphene edge, so that it no longer follows the same reaction pathway as the 4.5–7.0 Å series (see the SI). The lower limit of 4.5 Å corresponds to the extreme repulsive regime, beyond which the hexamer structure would become physically unreasonable.

In free space, the cyclic water hexamer adopts a puckered chair conformation in which each water molecule acts as both a proton donor and a proton acceptor, with the dangling O-H bonds alternating above and below the ring. To quantify the ring geometry, the ring plane is defined as the least-squares plane fitted to the six oxygen atoms, and the root-mean-square deviation of the oxygen atoms from this plane ($RMSD_O$, where the subscript O denotes oxygen) measures the out-of-plane puckering of the oxygen skeleton, as shown in Fig. 1(a). The reported bond lengths and angles are averaged over the six equivalent positions of the ring. For the equilibrium free-space structure, $RMSD_O$ is 0.144 Å, whereas the RMSD of all eighteen atoms ($RMSD_{all}$) reaches 0.547 Å because the dangling O-H bonds extend farther out of the plane. The O···O distance $d_{O\cdots O}$ between adjacent hydrogen-bonded oxygen atoms is 2.704 Å, and the corresponding H···O distance $d_{H\cdots O}$ is 1.723 Å. These free-space values serve as the reference against which the confinement-induced structural changes are compared in the following analysis. The complete set of structural parameters for both the equilibrium and transition-state structures (abbreviated as EQ and TS) at all interlayer spacings is collected in Table S1. Figure 1(b) schematically shows the confined geometry, and figure 1(c) illustrates the three stages of the concerted proton transfer (reactant, transition state, and product).

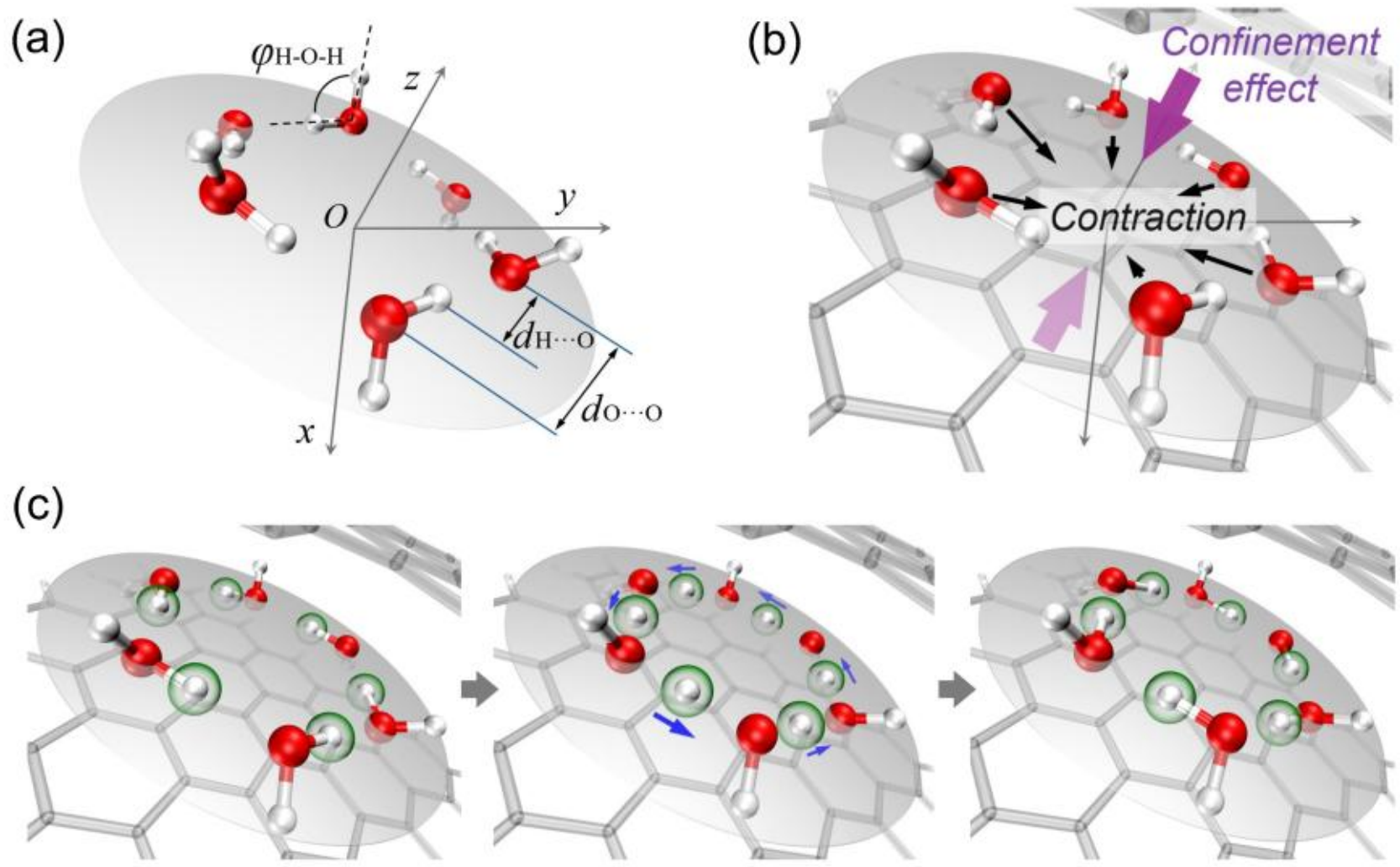

**Figure 1.** Oblique views of the water hexamer (a) in free space and (b) under confinement, and (c) the concerted proton-transfer process. (a) marks the O···O distance ($d_{O\cdots O}$), the H···O distance ($d_{H\cdots O}$), and the H-O-H angle ($\varphi_{H\text{-}O\text{-}H}$). (b) shows the hexamer compressed between two graphene sheets, with the arrows indicating the in-plane contraction of the ring. (c) shows the reactant, transition state, and product, with the six transferred protons circled. The gray translucent disk in all panels marks the ring plane, defined as the least-squares plane fitted to the six oxygen atoms.

Under confinement, the graphene walls act on the hexamer along the normal to its plane and progressively flatten the chair-like puckered ring that the hexamer adopts in free space. In free space the ring puckers to relieve the ring strain. This strain arises because a planar six-membered ring forces the O···O···O angles to 120°, whereas each water molecule prefers a tetrahedral arrangement of its hydrogen bonds (about 109.5°). However, this flattening is not monotonic as the interlayer spacing is reduced from 7.0 Å toward 4.5 Å (Fig. 2(a)). $RMSD_O$ first decreases from 0.092 Å to 0.026 Å at 6.5 Å, where the maximal face-on attraction outweighs the ring strain. It then rebounds to 0.075–0.074 Å at 6.0–5.5 Å as the attraction weakens and the ring strain reappears. Finally, it drops to 0.002 Å at 4.5 Å, where the interaction becomes strongly repulsive and compresses the ring almost perfectly into the plane. The minimum in $RMSD_O$ at 6.5 Å arises because this spacing places the hexamer at the van der Waals equilibrium distance (≈3.2 Å) from both walls at once. This maximal face-on attraction is reflected in the most negative water-graphene interaction energy $E_{int} = -1.251$ eV. This attraction pulls the oxygen atoms toward the ring plane, lowering $RMSD_O$ to 0.026 Å. At this spacing the dangling O-H bonds are not compressed by the walls and remain free. Consequently, the $RMSD_{all}$ keeps increasing even as the oxygen skeleton flattens. The flattening at 6.5 Å is therefore an active, adsorption-driven flattening. In contrast, at the tighter spacings of 4.5–5.0 Å the ring is flattened passively by the strong repulsion. Here $E_{int}$ is strongly repulsive (+6.887 and +1.627 eV), and the walls squeeze the ring into the plane.

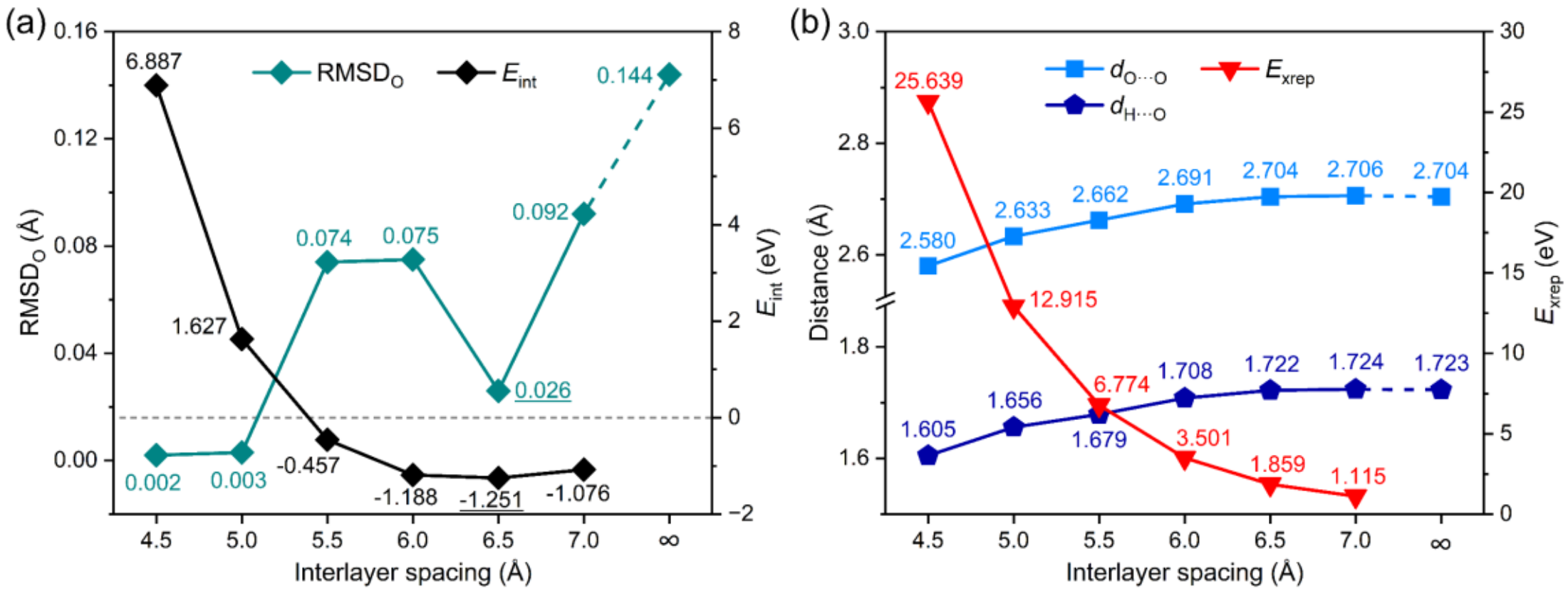

**Figure 2.** Structural parameters of the optimized equilibrium structures as functions of the interlayer spacing. (a) Root-mean-square deviation of the oxygen atoms from the least-squares plane, RMSD$_O$ (left axis) and the water-graphene interaction energy $E_{int}$ (right axis, ωB97X-D). (b) $d_{O\cdots O}$ and $d_{H\cdots O}$ (left axis) and the Pauli repulsion (exchange-repulsion) term $E_{xrep}$ (right axis, BHandHLYP-D3(BJ)).

Besides the flattening of the ring, confinement also compresses the hydrogen-bond network in the plane. As the interlayer spacing decreases from 7.0 to 4.5 Å, $d_{O\cdots O}$ shortens monotonically from 2.706 to 2.580 Å, and $d_{H\cdots O}$ shortens as well, from 1.724 to 1.605 Å (Fig. 2(b)). This in-plane contraction runs counter to the Poisson effect familiar in elastic materials, where a material compressed along one direction expands in the perpendicular directions (Fig. 1(b)). The contraction arises from the reorientation of the water molecules from a tetrahedral to a planar arrangement, rather than from a direct lateral push on the oxygen atoms. The Pauli repulsion from the walls presses the dangling O-H bonds and dangling lone pairs into the ring plane, driving this reorientation. The planarized molecules are then forced open by the 120° O···O···O angles of the ring, so the intramolecular H-O-H angle $\varphi_{H\text{-}O\text{-}H}$ widens from 105.92° in free space to 107.76° at 4.5 Å. This planarization also strengthens the hydrogen bonds, shortening $d_{O\cdots O}$ and $d_{H\cdots O}$. This monotonic contraction of the hydrogen-bond network is driven by the Pauli repulsion $E_{xrep}$ of the interaction, which increases with confinement from 1.115 eV at 7.0 Å to 25.639 eV at 4.5 Å, independent of the non-monotonic flattening.

These responses expose a clear separation between soft and stiff degrees of freedom of the ring. The soft degrees, the out-of-plane puckering and the hydrogen-bond network ($d_{O\cdots O}$, $d_{H\cdots O}$), deform readily under confinement and carry most of the imposed strain. By contrast, the stiff covalent O-H bond lengths, both donor and dangling, and the $\varphi_{H\text{-}O\text{-}H}$ angle change only modestly, the O-H lengths varying by only about 0.005 Å across the series. Confinement modulates the water hexamer through two soft geometric channels, namely the flattening of the ring and the compression of its hydrogen-bond network, both rooted in the wall-hexamer interaction. The transition-state structures display similar trends. Whether this geometric modulation affects the concerted proton-transfer barrier is examined in what follows.

Confinement lowers the barrier of the concerted proton transfer. The electronic barrier, computed at the DLPNO-CCSD(T) level, decreases monotonically as the confinement tightens, from 1.545 eV in free space to 1.139 eV at 4.5 Å, a reduction of 0.406 eV (Fig. 3(a)). The ωB97X-D barriers follow the same trend, confirming that the reduction is robust to the level of theory. This lower barrier does not arise from the structural changes shown in the preceding section. When the graphene walls are removed while the confined geometry is kept frozen, the intrinsic barrier $\Delta E^{\ddagger\prime}$ of the bare water hexamer remains essentially unchanged across the full O···O compression from 2.706 to 2.580 Å (the transition-state O···O distance stays nearly constant at 2.413–2.438 Å), fluctuating over a range of only 0.040 eV (1.534–1.574 eV) at the DLPNO-CCSD(T) level. The geometric changes are thus merely a consequence of confinement. Zero-point energy corrections, computed from the ωB97X-D frequencies, are largest in free space and decrease under confinement, but the tight-confinement barrier reduction remains (see the SI).

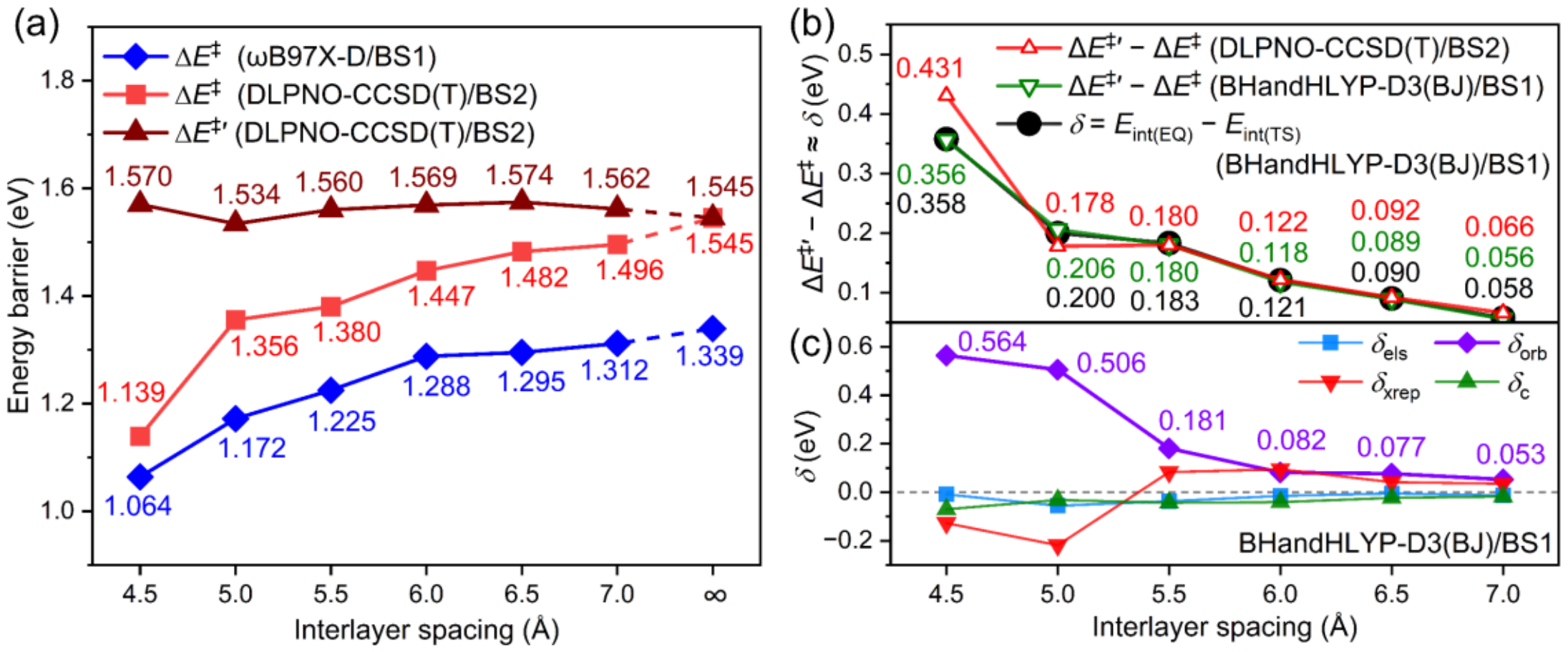

**Figure 3.** (a) Concerted proton-transfer barrier $\Delta E^{\ddagger}$ of the confined water hexamer at the ωB97X-D/BS1 and DLPNO-CCSD(T)/BS2 levels, and the intrinsic barrier $\Delta E^{\ddagger\prime}$ of the bare water hexamer frozen in its confined geometry (DLPNO-CCSD(T)/BS2), as a function of interlayer spacing. (b) The barrier reduction $\Delta E^{\ddagger\prime} - \Delta E^{\ddagger}$ (DLPNO-CCSD(T)/BS2 and BHandHLYP-D3(BJ)/BS1) and the differential interaction $\delta = E_{\mathrm{int(EQ)}} - E_{\mathrm{int(TS)}}$ at the BHandHLYP-D3(BJ)/BS1 level. At the same level, $\Delta E^{\ddagger\prime} - \Delta E^{\ddagger}$ nearly coincides with $\delta$, differing only by the small graphene-framework term, showing that the barrier reduction arises from the water-graphene electronic interaction. (c) Decomposition of $\delta$ into its sobEDA components $\delta_{\mathrm{els}}$, $\delta_{\mathrm{orb}}$, $\delta_{\mathrm{xrep}}$, and $\delta_{\mathrm{c}}$; positive values indicate a lowering of the transition-state energy.

The barrier reduction instead comes from the graphene lowering the transition-state energy more than the equilibrium structure. We define $\delta = E_{\mathrm{int(EQ)}} - E_{\mathrm{int(TS)}}$, the amount by which the transition state is adsorbed more strongly, or repelled less strongly, than the equilibrium structure. Within the activation strain model, the composite barrier decomposes as $\Delta E^{\ddagger} = \Delta E^{\ddagger\prime} + \Delta E_{\mathrm{graphene}} - \delta$, where $\Delta E_{\mathrm{graphene}} = E_{\mathrm{graphene(TS)}} - E_{\mathrm{graphene(EQ)}}$ is a small graphene-framework term. Rearranging, the barrier reduction is $\Delta E^{\ddagger\prime} - \Delta E^{\ddagger} = \delta - \Delta E_{\mathrm{graphene}}$, and because $\Delta E_{\mathrm{graphene}}$ is at most $6.5 \times 10^{-3}$ eV, the BHandHLYP-D3(BJ) curve of $\Delta E^{\ddagger\prime} - \Delta E^{\ddagger}$ coincides with $\delta$ to within this small term (Fig. 3(b)). The DLPNO-CCSD(T) curve of $\Delta E^{\ddagger\prime} - \Delta E^{\ddagger}$ closely follows the BHandHLYP-D3(BJ) curve across the series, the largest deviation (~0.07 eV at 4.5 Å) reflecting the difference between the two levels of theory. The BHandHLYP-D3(BJ) decomposition of $\delta$, which DLPNO-CCSD(T) cannot provide, therefore offers a reliable basis for interpreting the DLPNO-CCSD(T) barrier reduction. $\delta$ is positive at every spacing and grows monotonically with confinement, from 0.058 eV at 7.0 Å to 0.358 eV at 4.5 Å, so the graphene lowers the transition-state energy progressively more than that of the equilibrium structure as the confinement tightens. A sobEDA decomposition of $\delta$ singles out the orbital contribution $\delta_{orb}$, which rises from +0.053 eV at 7.0 Å to +0.564 eV at 4.5 Å and is the only contribution that is positive at every spacing (Fig. 3(c)). Because the orbital interaction is attractive ($E_{orb} < 0$), the positive $\delta_{orb} = E_{orb(EQ)} - E_{orb(TS)}$ means that the transition state enjoys a stronger, and hence more stabilizing orbital interaction than the equilibrium structure. The exchange-repulsion contribution $\delta_{xrep}$ changes sign across the series, assisting the barrier reduction in the weak-confinement regime (+0.035 eV at 7.0 Å) and opposing it in the strong-confinement regime (−0.127 eV at 4.5 Å). This sign change, together with its non-monotonic trend, shows that the monotonic barrier reduction is not driven by $\delta_{xrep}$, but is dominated by the always positive, monotonic orbital contribution $\delta_{orb}$, while $\delta_{xrep}$ partially offsets $\delta_{orb}$ in the strong-confinement regime. The electrostatic and correlation contributions $\delta_{els}$ and $\delta_c$ are small and slightly negative. Therefore, the barrier reduction is carried by $\delta$, dominated by the orbital stabilization of the transition state. This decomposition closes at both the BHandHLYP-D3(BJ) and B3LYP-D3(BJ) levels (see the SI).

The orbital stabilization has a clear structural origin. In the transition state the six protons sit approximately midway between their donor and acceptor oxygens, forming O···H···O three-center bonds, and the O···O distance contracts to 2.413–2.438 Å across the whole series. The water cluster in this state is more diffuse electronically and is polarized more readily by the graphene π face, which strengthens the orbital interaction. This gain outweighs the simultaneous rise of the Pauli repulsion as the cluster moves closer to the walls, so the transition-state energy is lowered overall. The equilibrium structure, by contrast, keeps its protons localized on the donor side, its electron density localized and hard to polarize, and its orbital interaction weak. ETS-NOCV analysis identifies two channels, the donation from the water lone pairs into the graphene π* orbitals and the back-donation from the graphene π orbitals into the water σ* orbitals. The donation dominates. At 4.5 Å its summed energy (−1.802 eV in the transition state) exceeds the back-donation (−0.079 eV) by more than an order of magnitude (see the SI). The donation energy grows from −1.519 eV in the equilibrium structure, in line with the positive $\delta_{orb}$. Figure 4 visualizes these channels. The donation shows electron density depleted around the water ring (red) and accumulated in the interfacial region (blue), more concentrated in the transition state. The back-donation is far weaker. Confinement therefore lowers the barrier through one electronic channel, the orbital stabilization of the transition state, which runs in parallel with, and independent of, the geometric modulation of the ring.

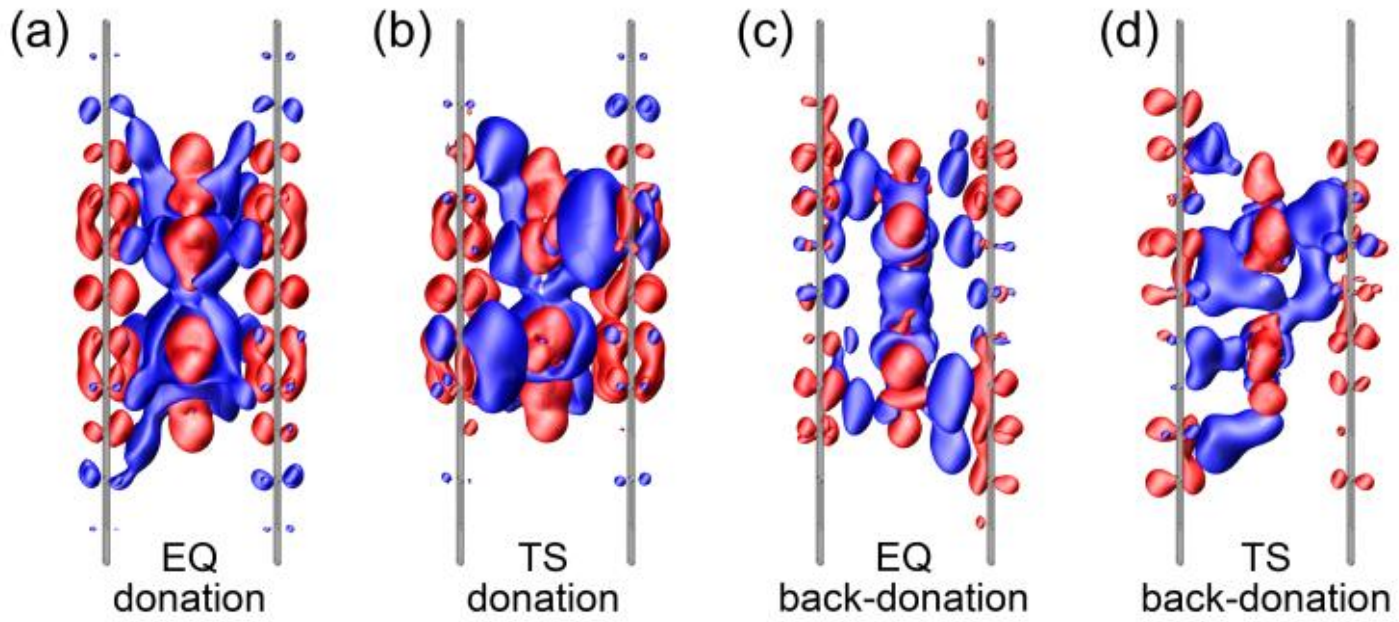

**Figure 4.** ETS-NOCV deformation densities of the donation (a, b) and back-donation (c, d) charge-transfer channels between the water cluster and the graphene walls in the confined hexamer at 4.5 Å, at the BHandHLYP-D3(BJ)/BS1 level, for the equilibrium structure (a, c) and the transition state (b, d). Blue and red isosurfaces denote electron density accumulation and depletion, respectively. Isosurface values: ±0.0005 e/Bohr³ for donation and ±0.0001 e/Bohr³ for back-donation.

**Table 1.** Transition-state imaginary frequency, O-H stretching frequencies, and proton-transfer barrier of the confined water hexamer.[a]

| Interlayer Spacing (Å) | $\lvert\nu(\mathrm{TS})\rvert$ (cm$^{-1}$) | O-H$^{donor}$ (cm$^{-1}$) | O-H$^{dangling}$ (cm$^{-1}$) | $\Delta E^{\ddagger}$ (eV) |
|---|---|---|---|---|
| 4.5 | 430 | 3418 | 3991 | 1.139 |
| 5.0 | 501 | 3462[b] | 3967 | 1.356 |
| 5.5 | 725 | 3419 | 3935 | 1.380 |
| 6.0 | 884 | 3431 | 3915 | 1.447 |
| 6.5 | 864 | 3448 | 3907 | 1.482 |
| 7.0 | 927 | 3457 | 3911 | 1.496 |
| ∞ | 1652 | 3469 | 3946 | 1.545 |

[a] Frequencies were computed at the ωB97X-D/BS1 level (BS1: def2-TZVP for water and def2-SVP for graphene); barriers were computed at the DLPNO-CCSD(T)/BS2 level (BS2: def2-QZVPP for water and def2-SVP for graphene).

[b] The donor O-H frequency at 5.0 Å deviates from the trend owing to a numerical difficulty in the transition region rather than a physical effect; see the SI for details.

The vibrational spectrum offers a complementary view of this physics. The magnitude of the imaginary frequency of the transition state, which sets the curvature of the proton-transfer barrier, falls from 1652 cm$^{-1}$ in free space to 430 cm$^{-1}$ at 4.5 Å, so the barrier flattens broadly in parallel with its reduction from 1.545 to 1.139 eV (Table 1). The O-H stretching region resolves the two kinds of O-H bond. The donor O-H stretches soften slightly as the hydrogen bonds strengthen, whereas the dangling O-H stretches red-shift from free space to about 6.5 Å, where they are drawn toward the graphene, and then blue-shift toward 4.5 Å, where Pauli repulsion stiffens them. Consistent with this softening, the H···O Wiberg bond index grows from 0.061 in free space to 0.090 at 4.5 Å, while the donor O-H bond index falls from 0.683 to 0.611, with the electron density of the covalent O-H bond shifting into the hydrogen-bond contact. The Hirshfeld charge of the water cluster likewise rises monotonically from −0.152 e at 7.0 Å to +0.129 e at 4.5 Å, turning from negative to positive as the confinement tightens. The donor-acceptor interaction between the water lone pairs and the graphene π* orbitals grows from zero at 7.0 Å to 1.603 eV at 4.5 Å. The increasingly positive charge and the growing donation energy together mark a water-to-graphene charge transfer that strengthens with confinement. This charge transfer is part of the orbital stabilization captured by $\delta_{orb}$.

In summary, confinement lowers the electronic concerted proton-transfer barrier by 0.406 eV, from 1.545 eV in free space to 1.139 eV at 4.5 Å. The geometric modulation, the flattening of the ring and the compression of its hydrogen-bond network, is a consequence of the wall-hexamer interaction rather than the cause of this reduction. The barrier instead falls because the graphene π face lowers the transition-state energy through orbital donation from the water lone pairs. This mechanism strengthens monotonically with confinement and is corroborated by the imaginary frequency, the bond orders, and the charge measures, which together distinguish the electronic cause from the geometric consequence. The result identifies the electronic interaction with the confining walls, rather than the confinement geometry itself, as the factor that controls proton transfer, providing a mechanistic design principle for tailoring proton conduction and chemical reactivity through nanoscale confinement, relevant to nanofluidic devices and proton-exchange membranes.

## ■ COMPUTATIONAL METHODS

Geometry optimizations of the equilibrium structures and transition states, frequency calculations, and IRC calculations were performed using the long-range corrected hybrid functional ωB97X-D[28] in combination with the def2-TZVP basis set for water and def2-SVP for graphene (denoted BS1), a choice justified by the need for an accurate description of the water cluster while keeping the entire confinement system computationally feasible, with the Gaussian 16 program (Revision C.01).[29] The functional was selected by benchmarking the concerted proton-transfer barriers of cyclic water clusters $(H_2O)_n$ against reference values. For $n$ = 3–5, the CCSD(T)/CBS barriers of Xue et al.[30] served as the reference; for the hexamer, for which a CCSD(T)/CBS reference is not available in the literature, the B3LYP/aug-cc-pVDZ barrier of Elahi and Zeinalipour-Yazdi[31] was used as a lower-level reference. Both PBE0-D3(BJ) and ωB97X-D underestimate these barriers, but PBE0-D3(BJ) by 0.326–0.480 eV, far more than ωB97X-D (0.085–0.166 eV). After DLPNO-CCSD(T) single-point correction, the ωB97X-D geometry reproduces the CCSD(T)/CBS barriers ($n$ = 3–5) to within 0.004 eV, closer than the PBE0-D3(BJ) geometry (within 0.021 eV). We therefore optimize all structures with ωB97X-D and refine the barriers with DLPNO-CCSD(T) (see the SI for the full comparison). For each confinement condition, geometry optimizations were performed to obtain the equilibrium structures of the cyclic water hexamer, with the graphene atoms held fixed, and the transition state for concerted proton transfer was optimized using the Berny algorithm. Harmonic vibrational frequency analysis confirmed that each transition state possesses exactly one imaginary frequency, and intrinsic reaction coordinate (IRC) calculations verified that it connects the intended reactant and product. Vibrational frequencies and Raman activities were computed at the same ωB97X-D level, and Wiberg bond indices and second-order perturbation energies were obtained from natural bond orbital (NBO) analysis[32] together with Hirshfeld charges from Hirshfeld population analysis.

To obtain accurate barriers, single-point energies were computed at the DLPNO-CCSD(T) level[33] with the def2-QZVPP basis set for water and def2-SVP for graphene (denoted BS2; def2-QZVPP/C auxiliary basis) on the ωB97X-D geometries, with the ORCA program (version 6.1.0).[34] Because the graphene sheets serve only as a confining environment and the electron correlation of the graphene and the water-graphene interface barely changes along the reaction coordinate, a multilayer scheme was adopted.[35] The electron pairs within the graphene sheets and across the water-graphene interface were frozen at the Hartree-Fock level, while tight-PNO DLPNO-CCSD(T) correlation was applied only within the water hexamer. The resulting barrier sequence follows the same trend as the ωB97X-D barriers.

The ωB97X-D interaction energies were computed in the counterpoise convention (ghost basis functions) to remove the basis-set superposition error.[36] The interaction of the water hexamer with the graphene walls was decomposed by sobEDA energy decomposition analysis,[37] carried out with the Multiwfn program (version 3.8).[38] Since this decomposition is not available for long-range corrected functionals such as ωB97X-D currently, it is reported at the BHandHLYP-D3(BJ) level with BS1, with all fragment energies evaluated in the same counterpoise convention. The higher Hartree-Fock exchange in BHandHLYP-D3(BJ) (50% versus 20% in B3LYP-D3(BJ)) captures the exchange-repulsion term $E_{xrep}$, the dominant term in the structural change of the water ring, more faithfully. The orbital term $E_{orb}$, central to the analysis of the concerted proton-transfer barrier, differs little between the two functionals, and the B3LYP-D3(BJ) results follow the same trends and are given in the SI. In the sobEDA decomposition, the exchange ($E_x$) and orthogonalization-repulsion ($E_{rep}$) contributions arise jointly from wavefunction antisymmetrization, and we refer to their sum $E_{xrep} = E_x + E_{rep}$ as the Pauli repulsion throughout this work. All terms are evaluated with the same functional so that they sum to the total interaction $E_{int}$, as required for a closed decomposition. To disentangle the geometric and electronic contributions to the barrier, the barrier was decomposed within the activation strain model (also called the distortion/interaction model).[39, 40] In this model, the composite barrier $\Delta E^{\ddagger}$ is decomposed into the intrinsic water barrier $\Delta E^{\ddagger\prime}$, the concerted proton-transfer barrier of the bare water hexamer frozen in its confined geometry, the differential interaction $\delta = E_{int(EQ)} - E_{int(TS)}$, and a small graphene-framework term $\Delta E_{graphene} = E_{graphene(TS)} - E_{graphene(EQ)}$, such that $\Delta E^{\ddagger} = \Delta E^{\ddagger\prime} + \Delta E_{graphene} - \delta$. Because $\Delta E_{graphene}$ is small ($\leq 6.5 \times 10^{-3}$ eV), $\Delta E^{\ddagger\prime} - \Delta E^{\ddagger}$ nearly coincides with δ, as discussed in the main text. The intrinsic barrier $\Delta E^{\ddagger\prime}$, which serves to test whether the structural change of the ring affects the barrier, is evaluated at the DLPNO-CCSD(T) level for the main text and at the BHandHLYP-D3(BJ) and B3LYP-D3(BJ) levels for the closed decomposition, in each case with the graphene ghost basis functions retained (see the SI for the resulting values).

## ■ ASSOCIATED CONTENT

### Supporting Information

The Supporting Information is available free of charge at https://xxx.xxx.xxx/doi/10.xxxx/xxxxxxx

The Supporting Information consists of seven sections: S1. Equilibrium and transition-state geometries; S2. Method validation on cyclic water clusters; S3. Proton-transfer barriers; S4. Energy decomposition and closure; S5. ETS-NOCV deformation densities; S6. Vibrational, bond-order, and charge evidence; S7. Reliability and boundary evidence. It contains Figures 1–3 and Tables 1–13. (PDF)

The data that support the findings of this study are available from the corresponding author upon reasonable request.

## ■ AUTHOR INFORMATION

### Corresponding Author

**Yang Gao** − *School of Medicine, Southwest University of Science and Technology, Mianyang 621010, China;* orcid.org/0000-0001-9148-8635

Email: ygaoxs@gmail.com

### Authors

**Le Jin** − *School of Science, Shenyang University, Shenyang 110044, China;* orcid.org/0009-0004-3696-2911

**Depeng Zhang**− *School of Science, Shenyang University, Shenyang 110044, China;* orcid.org/0000-0002-2483-0573

**Xinrui Yang** − *Key Laboratory of Theoretical and Computational Photochemistry, Ministry of Education, College of Chemistry, Beijing Normal University, Beijing 100875, China;* orcid.org/0000-0002-9234-1309

**Xiaoan Li** − *NHC Key Laboratory of Nuclear Technology Medical Transformation, Mianyang Central Hospital, School of Medicine, University of Electronic Science and Technology of China, Mianyang, Sichuan, 621099, China; Department of Gastroenterology, National Clinical Key Specialty, Mianyang Central Hospital, School of Medicine, University of Electronic Science and Technology of China, Mianyang 621000, China;* orcid.org/0009-0006-8293-8845

**Bo Wang** − *School of Science, Northeast Electric Power University, Jilin 132012, China*

Complete contact information is available at: http://xxx.xxx.xxx/10.xxxx/xxx.xxxxxxx.xxxxxxx

### Author Contributions

**Le Jin:** Data curation (lead); Investigation (lead); Formal analysis (lead); Methodology (equal); Project administration (supporting); Writing – original draft (lead); Writing – review & editing (lead); **Depeng Zhang:** Conceptualization (lead); Methodology (equal); Funding acquisition (equal); Project administration (equal); Resources (equal); Supervision (equal); Writing – original draft (equal); Writing – review & editing (equal); **Xinrui Yang:** Conceptualization (supporting); Formal analysis (equal); Investigation (supporting); Validation (supporting); Writing – review & editing (supporting); **Xiaoan Li**: Investigation (supporting); Formal analysis (supporting); Validation (supporting);

Writing – review & editing (supporting); **Bo Wang:** Conceptualization (supporting); Investigation (supporting); Writing – review & editing (supporting); **Yang Gao:** Conceptualization (lead); Methodology (lead); Funding acquisition (lead); Project administration (lead); Resources (lead); Supervision (lead); Writing – original draft (equal); Writing – review & editing (equal); Data curation (equal); Formal analysis (equal); Investigation (equal); Validation (equal); Visualization (equal).

## Notes

The authors declare no competing financial interest.

## ■ ACKNOWLEDGMENTS

This work was supported by the NHC Key Laboratory of Nuclear Technology Medical Transformation (Mianyang Central Hospital) (Grant No.2022HYX002), and the Tianfu Jiangxi Laboratory (No. TFJX-ZD-2024-005), and the Basic Scientific Research Project of Liaoning Provincial Department of Education (Grant No. LJ212511035008).

## ■ REFERENCES